\documentclass[aps,pre,twocolumn,superscriptaddress,nofootinbib]{revtex4-1}
\usepackage{graphicx}
\usepackage{dcolumn}
\usepackage{bm}
\usepackage{amsmath} 
\usepackage{lipsum} 
\usepackage{hyperref}
\hypersetup{colorlinks=true,linkcolor=blue,citecolor=blue,urlcolor=blue}
\usepackage[normalem]{ulem}

\begin{document}

\title{Game-environment feedback dynamics for voluntary prisoner's dilemma games}
\author{Bin-Quan Li}
\author{Cong Liu}
\author{Zhi-Xi Wu}\email{wuzhx@lzu.edu.cn}
\author{Jian-Yue Guan}
\affiliation{Lanzhou Center for Theoretical Physics and Key Laboratory of Theoretical Physics of Gansu Province, Lanzhou University, Lanzhou, Gansu 730000, China}
\affiliation{Institute of Computational Physics and Complex Systems, Lanzhou University, Lanzhou, Gansu 730000, China}

\begin{abstract}
  Recently, the eco-evolutionary game theory which describes the coupled dynamics of strategies and environment have attracted great attention. At the same time, most of the current work is focused on the classic two-player two-strategy game.
  In this work, we study multi-strategy eco-evolutionary game theory which is an extension of the framework. For simplicity, we'll focus on the voluntary participation Prisoner's dilemma game.
  For the general class of payoff-dependent feedback dynamics, we show the conditions for the existence and stability of internal equilibrium by using the replicator dynamics, respectively.
  Where internal equilibrium points, such as, two-strategy coexistence states, three-strategy coexistence states, persistent oscillation states and interior saddle points.
  These states are determined by the relative feedback strength and payoff matrix, and are independent of the relative feedback speed and initial state.
  In particular, the three-strategy coexistence provides a new mechanism for maintaining biodiversity in biology, ecology, and sociology.
  Besides, we find that this three-strategy model return to the persistent oscillation state of the two-strategy model when there is no defective strategy at the initial moment.
\end{abstract}
\maketitle

\section{Introduction}

Cooperative behavior is ubiquitous in real systems, such as biological systems, social systems and economic systems \cite{Gore2009,Lee2010,Lewin-Epstein2017}.
Cooperative behavior plays a crucial role in the normal operation of these systems.
Over the past few decades, evolutionary game theory has been introduced into the study of cooperative phenomena, and it is found that this theoretical framework is very effective in dealing with this problem \cite{Nowak2006,Taylor2007,Nowak2004,Lleberman2005,Ohtsuki2006,Chen2021}.
In well-mixed infinite populations, dynamics are usually described by the replicator equation \cite{Taylor1978,Hofbauer1979,Nowak2006b}.
In the class evolutionary game theory, a cooperator ($C$) helps all individuals to whom it is connected.
A defector ($D$) does not provide any help, and therefore has no costs, but it can receive the benefit from neighboring cooperators.
It is worth noting that defectors often threaten the success of the common cause as they try to free ride from the efforts of the community.

The Prisoner's Dilemma game (PDG) was introduced to study the emergence and maintenance of cooperation between selfish individuals in society \cite{Poundstone1992,Szabo1998,Wu2005,Wu2006,Wu2007,Dai2011,Fu2021}.
Mutual defection is a Nash equilibrium for the PDG.
Therefore, the defection strategy is the optimal choice and does not depend on the opponent's strategy.
It is well known that natural selection favors defectors over cooperators in unstructured populations.
Defectors have a higher average payoff than cooperators in the PDG.
Therefore, natural selection increases the relative abundance of defectors and drives cooperators to extinction.
The loner ($L$) strategy was introduced to prevent the uniform defection \cite{Hauert2002,Hauert2002S}.
This strategy represents players who wish to avoid the risk of being exploited.
For this purpose, they refuse to participate in the game and are content to share their lower income with their playmates.
Thus, loners can foil defectors and overcome a social dilemma.
The effects of loners have been widely analyzed using public goods games \cite{Szabo2002PRL,Semmann2003,Hu2019} and PDG \cite{Szabo2002,Szabo2004,Chu2017,Cardinot2018,Yamamoto2019}.
These three strategies lead to a ``rock-paper-scissors" dynamics with cyclic dominance ($L$ invades $D$ invades $C$ invades $L$) \cite{Christian2008,Szolnoki2014,Cangiani2018,Guo2020,Park2020}.

A lot of previous studies have shown that the environment is fixed and does not change over time.
Real-world systems, however, often feature bi-directional feedbacks between the environment and the incentives in strategic interactions.
From microbe to human society, strategy-dependent feedback is widely existed.
For example, overgrazing will lead to grassland degradation, and grazing control will make grassland recovery. In fisheries, the yield depends on the biomass of the fish; In turn, the stock biomass depends on the frequency of the fishing strategy.

Recently, more and more people pay attention to the interaction between collective environment and individual behavior \cite{Wardil2013,Hilbe2018,Su2019a,Mao2021,Luo2021}.
A new theoretical framework has been proposed, which further describes the coupled evolution of strategies and the environment \cite{Weitz2016,Lin2019,Tilman2020}.
Weitz \emph{et al.} \cite{Weitz2016} describe an oscillatory tragedy of the commons in which the system cycles between deplete and replete environmental states and cooperation and defection behavior states.
They found that the conditions in the model to avoid the tragedy of the commons depended on the strength of the coupling, not the speed of the coupling.
That is the qualitative dynamics remain invariant with different relative feedback speed.
Based on this framework, Tilman \emph{et al.} \cite{Tilman2020} proposed a more general framework of eco-evolutionary games that consider environments controlled by internal growth, decline, or tipping points. And they found that the dynamical behaviors actually largely depend on the relative feedback speed.

However, much of the current work is focused on the classic two-player two-strategy game \cite{Wu2019,Glaubitz2020,Pastor2020,Wang2020,Yang2021,Cao2021,Yan2021,Bairagya2021}. There are few studies on environmental feedback dynamics with three or more strategies.
Based on this, we can naturally extend the analysis of the evolutionary dynamics of multi-strategy games with environmental feedback, for example, the voluntary participation Prisoner's dilemma game.

In this paper, we extend the classical two-strategy model to three-strategy model in the well-mixed infinite population. For the general reward-dependent feedback dynamics, we will use the replicator dynamics to explore the conditions for the existence and stability of the internal equilibrium of the system.

\section{Model}
The voluntary prisoner's dilemma has been well studied \cite{Szabo2002,Szabo2004,Chu2017,Cardinot2018,Yamamoto2019}, and the general payoff matrix is as follows
\begin{equation}
\label{eq.1}
  \bordermatrix{
    & C & D & L \cr
  C & b-c & -c & \delta \cr
  D & b & 0 & \delta \cr
  L & \delta & \delta & \delta \cr
  }.
\end{equation}
The payoff matrix represents two players choosing from three options of actions:
cooperation $C$, defection $D$, or loner $L$. If both players are cooperators, the payoff will be $b-c$, where $b>c>0$. If one player is defector and the other is cooperator, the former gets $b$, the latter gets $c$. If it was mutual defectors, nothing happened. If one player chooses the loner strategy, both players are rewarded $\delta$ ($0 < \delta < 1$).

We begin from the introduction of a generalized linear environment-dependent payoff structure which the strategies and the environment co-evolve is proposed by Weitz \emph{et al.} \cite{Weitz2016}, which can be represented as
\begin{equation}
\label{eq.2}
  A(n)=(1-n)\left[
  \begin{array}{ccc}
    \delta  & \delta & \delta \\
    b & 0 & \delta \\
    b-c & -c & \delta \\
  \end{array}
\right]
+n\left[
  \begin{array}{ccc}
    b-c & -c & \delta \\
    b & 0 & \delta \\
    \delta  & \delta & \delta \\
  \end{array}
\right],
\end{equation}
or, alternatively
\begin{equation}
\label{eq.3}
A(n)=\left[
\begin{array}{ccc}
\delta+n(b-c-\delta)  & \delta-n(c+\delta) & \delta \\
b & 0 & \delta \\
b-c+n(\delta+c-b) & n(c+\delta)-c & \delta \\
\end{array}
\right],
\end{equation}
where $0 \leq n \leq 1$ denotes the current state of the environment. $n=0$ and $n=1$ indicate that the environment state is depleted and replete, respectively.
The larger $n$ denotes a richer environment.

Thus, the fitness of cooperators, defectors and loners, denoted as $r_{C}$, $r_{D}$ and $r_{L}$, can be calculated as
\begin{equation}
\label{eq.4}
\begin{split}
	& r_{C}=x \left[ \delta+n(b-c-\delta) \right] + y \left[ \delta-n(c+\delta) \right]+z\delta , \\
	& r_{D}=x b + z \delta , \\
	& r_{L}=x \left[ b-c+n(\delta+c-b) \right] + y \left[ n(c+\delta)-c \right]+z\delta,
\end{split}
\end{equation}
where $x$, $y$, and $z$ represent the frequency of cooperators, defectors, and loners in the population, respectively.
We set $x+y+z=1$, so the independent variables become $x$ and $y$.

The standard replicator dynamics for the fraction of cooperators $x$ and the fraction of defectors $y$ are
\begin{equation}
\label{eq.5}
\left\{
  \begin{array}{ll}
   \dot{x}=x \left( r_{C}-\overline{r} \right) , \\
   \dot{y}=y \left( r_{D}-\overline{r} \right) , \\
   \dot{z}=z \left( r_{L}-\overline{r} \right) ,
  \end{array}
\right.
\end{equation}
where $\overline{r}$ represents the average fitness of the system as follows:
\begin{equation}
\label{eq.6}
 \overline{r}=x r_{C}+y r_{D}+z r_{L}.
\end{equation}
Meanwhile, the environment is modified by the population strategy states, and the environmental evolution is described by
\begin{equation}
\label{eq.7}
   \dot{n}=\varepsilon n(1-n)f(x,y,z),
\end{equation}
where $\varepsilon$ denotes the relative speed by which individual actions modify the environmental state and the term $n(1-n)$ ensures that the environmental state is confined to the domain $[0,1]$.
In addition, $f(x,y,z)$ denotes the feedback of strategists with the environment and a simple linear feedback mechanism is adopted:
\begin{equation}
\label{eq.8}
 f(x,y,z)=\theta x-y-z,
\end{equation}
where $\theta > 0$ indicates the relative strength of cooperators in enhancing the environment. In this simple scenario, environment can become better if there are more cooperators.

\section{Results and discussion}
By solving the equations $(\dot{x}=0,\dot{y}=0,\dot{n}=0)$, that is
\begin{equation}
\label{eq.9}
\left\{
  \begin{array}{ll}
   0=x \left( r_{C}-\overline{r} \right) , \\
   0=y \left( r_{D}-\overline{r} \right) , \\
   0=\varepsilon n(1-n)f(x,y,z).
  \end{array}
\right.
\end{equation}
We get ten positive equilibrium solutions. Of these, six represent ``boundary" fixed points:

(\romannumeral1) $\left(x^{*}=0,y^{*}=0,n^{*}=0\right)$, loners in a degraded environment;

(\romannumeral2) $\left(x^{*}=0,y^{*}=0,n^{*}=1\right)$, loners in a replete environment;

(\romannumeral3) $\left(x^{*}=0,y^{*}=1,n^{*}=0\right)$, defectors in a degraded environment;

(\romannumeral4) $\left(x^{*}=1,y^{*}=0,n^{*}=0\right)$, cooperators in a degraded environment;

(\romannumeral5) $\left(x^{*}=0,y^{*}=1,n^{*}=1\right)$, defectors in a replete environment, and

(\romannumeral6) $\left(x^{*}=1,y^{*}=0,n^{*}=1\right)$, cooperators in a replete environment.

There are also four interior fixed points:

(\romannumeral7) $\left(x^{*}=\frac{\delta}{b},y^{*}=1-\frac{\delta}{b},n^{*}=0\right)$, representing a mixed population of cooperators and defectors in a degraded environment;

(\romannumeral8) $\left(x^{*}=\frac{1}{1+\theta},y^{*}=0,n^{*}=\frac{1}{2}\right)$, representing a mixed population of cooperators and loners in an intermediate environment;

(\romannumeral9) $\left(x^{*}=\frac{1}{1+\theta},y^{*}=\frac{b+c-\delta}{(1+\theta)(\delta-c)},n^{*}=\frac{1}{2}\right)$, representing a mixed population of cooperators, defectors and loners in an intermediate environment, and

(\romannumeral10) $\left(x^{*}=\frac{1}{1+\theta},y^{*}=\frac{\theta}{1+\theta},n^{*}=\frac{\delta+\theta \delta-b}{(1+\theta)(\delta+c)-b}\right)$, representing a mixed population of cooperators and defectors in a high consumption environment.

Next we analyze the stability of the above four internal fixed points respectively.
The detailed theoretical analysis for the existence conditions and the stability of these fixed points are provided in \textbf{Appendix}.

\subsection{Stability conditions of internal equilibria}
\textbf{Case 1:} If $0 < \delta < \frac{1}{1+\theta}$, the internal equilibria $\left(\frac{\delta}{b},1-\frac{\delta}{b},0\right)$ is stable.

In this \textbf{Case 1}, when the evolution of the system is stable, cooperators and defectors occupy the whole system, loners become extinct and the environment state is depleted ($n=0$).

\begin{figure}[ht]
  \centering
  \includegraphics[width=0.47\textwidth,clip=false]{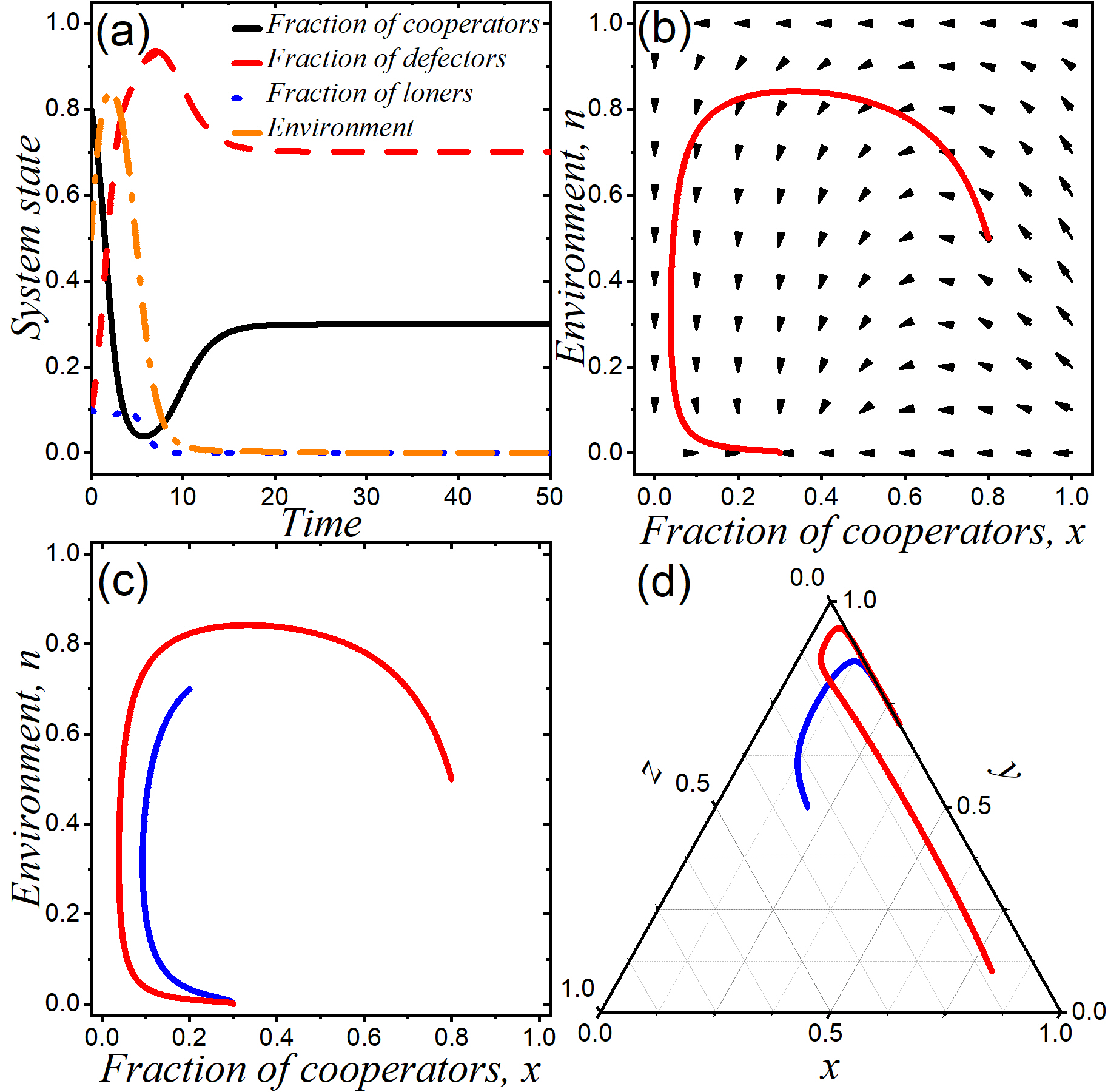}\\
  \caption{(Color online) (a) Time series of the fraction of cooperators $x$ (black), defectors $y$ (red), loners $z$ (blue) and the environmental state $n$ (orange), correspond to \textbf{Case 1} with $\varepsilon = 0.4$, $\theta = 2$, $b = 1$, $c = 0.6$ and $\delta = 0.3$.
  (b) Phase plane dynamics of $x-n$ system.  The red curve denotes the realized orbit. The black arrows denote the manifolds.
  The initial state is (0.8, 0.1, 0.5) and the final stable state is $(\frac{\delta}{b},1-\frac{\delta}{b},0)$.
  (c) Phase plane dynamics of $x-n$ system. (d) Phase plane dynamics of $x-y-z$ system.
  The blue and red curves represent trajectories with initial values of (0.2, 0.5, 0.7) and (0.8, 0.1, 0.3), respectively.	}
  \label{fig.1}
\end{figure}
For ease of understanding, a detailed example is provided in Fig.~\ref{fig.1}.
From the black manifold arrow, the red reality trajectory is counterclockwise in the $x-n$ phase plane as shown in Fig.~\ref{fig.1}(b).

The intuition is as follows.
At the initial moment, the population strategies state $x=0.8$ was relatively high in the system. As cooperators enhance the environment, the environmental state began to rise. Then, defectors will invade an environmentally enhanced state and the population strategies state $x$ began to decline. And then, in an environment dominated by defectors, the environmental state will be degraded.
Finally, as cooperators are favored in a degraded environment, and the system will be driven closer to ($\delta$, 0).
This intuition holds throughout the domain and different initial conditions are shown in Fig.~\ref{fig.1}(c).

Loners eventually go extinct, as can be seen from in Fig.~\ref{fig.1}(a) and (d).
This is completely different from the situation without environmental feedback. In the absence of environmental feedback, the system eventually stabilizes to be full of loners, while cooperators and defectors go extinct.

Due to $\delta < \frac{1}{1+\theta}$, the interaction between loneliness strategy and environment is very small, and eventually loneliness strategy will soon go extinct. Finally, cooperators and defectors coexist in the system. The frequency of cooperators and defectors is $\delta$ and $1-\delta$, respectively.

\textbf{Case 2:} If $\frac{1}{1+\theta} < \delta < \frac{1}{1+\theta}+c$, the internal equilibria $\left(\frac{1}{1+\theta},\frac{\theta}{1+\theta},\frac{\delta+\theta \delta-b}{(1+\theta)(\delta+c)-b}\right)$ is stable.

In this \textbf{Case 2}, when the evolution of the system is stable, cooperators and defectors occupy the whole system, loners become extinct, and the environmental state is a state of high consumption ($n<0.5$).

\begin{figure}[ht]
  \centering
  \includegraphics[width=0.47\textwidth,clip=false]{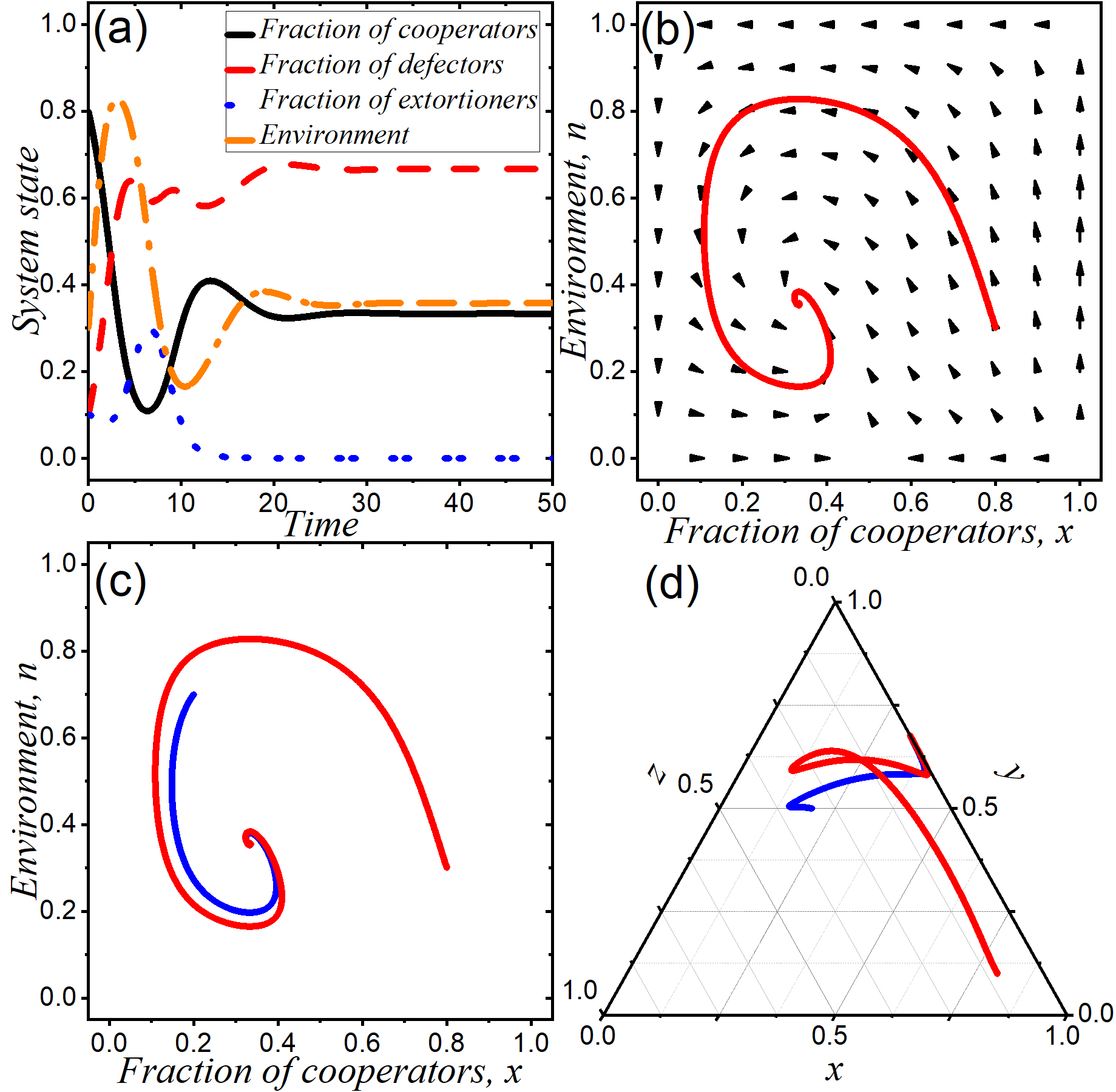}\\
  \caption{(Color online) (a) Time series of the fraction of cooperators $x$ (black), defectors $y$ (red), loners $z$ (blue) and the environmental state $n$ (orange), correspond to \textbf{Case 2} with $\varepsilon = 0.3$, $\theta = 2$, $b = 1$, $c = 0.3$ and $\delta = 0.5$.
  (b) Phase plane dynamics of $x-n$ system.  The red curve denotes the realized orbit. The black arrows denote the manifolds. The initial state is (0.8, 0.1, 0.3) and the final stable state is $(\frac{1}{1+\theta},\frac{\theta}{1+\theta},\frac{\delta+\theta \delta-b}{(1+\theta)(\delta+c)-b})$.
  (c) Phase plane dynamics of $x-n$ system. (d) Phase plane dynamics of $x-y-z$ system.
  The blue and red curves represent trajectories with initial values of (0.2, 0.5, 0.7) and (0.8, 0.1, 0.3), respectively.	}
  \label{fig.2}
\end{figure}

Figure~\ref{fig.2} shows an example corresponding to \textbf{Case 2}.
However, unlike \textbf{Case 1} above, the environment state is not a state of depleted ($n=0$), but a state of high consumption ($n<0.5$), and the environment state ends up at $\frac{\delta+\theta \delta-1}{(1+\theta)(\delta+c)-1}$.
As can be seen from $x-n$ phase diagram in Fig.~\ref{fig.2}(b), the system spirals counterclockwise towards the final stable state.

Due to $\frac{1}{1+\theta} < \delta < \frac{1}{1+\theta} +c$, the interaction between loneliness strategy and environment becomes greater.
Then, loners will invade an environment dominated by defectors, and the fraction of loners began to increase.
The biggest benefit brought by the increased proportion of loners is that cooperators can be further expanded.
As cooperators enhance the environment, the environmental state began to rise.
Although loners eventually went extinct, the state of the environment improved.
Finally, cooperators and defectors coexist in the system. The frequency of cooperators and defectors is $\delta$ and $1-\delta$, respectively, and the environment is no longer in a depletion state but in a high consumption state.

For different initial states, the final stable state of the system is consistent, as shown in Fig.~\ref{fig.2}(c) and (d).
The blue and red curves represent trajectories with initial values of $(0.2,0.5,0.7)$ and $(0.8,0.1,0.3)$, respectively.
It is clear from Fig.\ref{fig.2}(d) that the third strategy, loneliness, is ultimately extinct.

\textbf{Case 3:} If $\frac{1}{1+\theta}+c < \delta <1$, the internal equilibria $\left(\frac{1}{1+\theta},\frac{b+c-\delta}{(1+\theta)(\delta-c)},\frac{1}{2}\right)$ is stable.

In this \textbf{Case 3}, when the evolution of the system is stable, the three strategies of cooperation, defection and loneliness coexist, and the environment state is an intermediate state between replete and depleted.

\begin{figure}[ht]
  \centering
  \includegraphics[width=0.47\textwidth,clip=false]{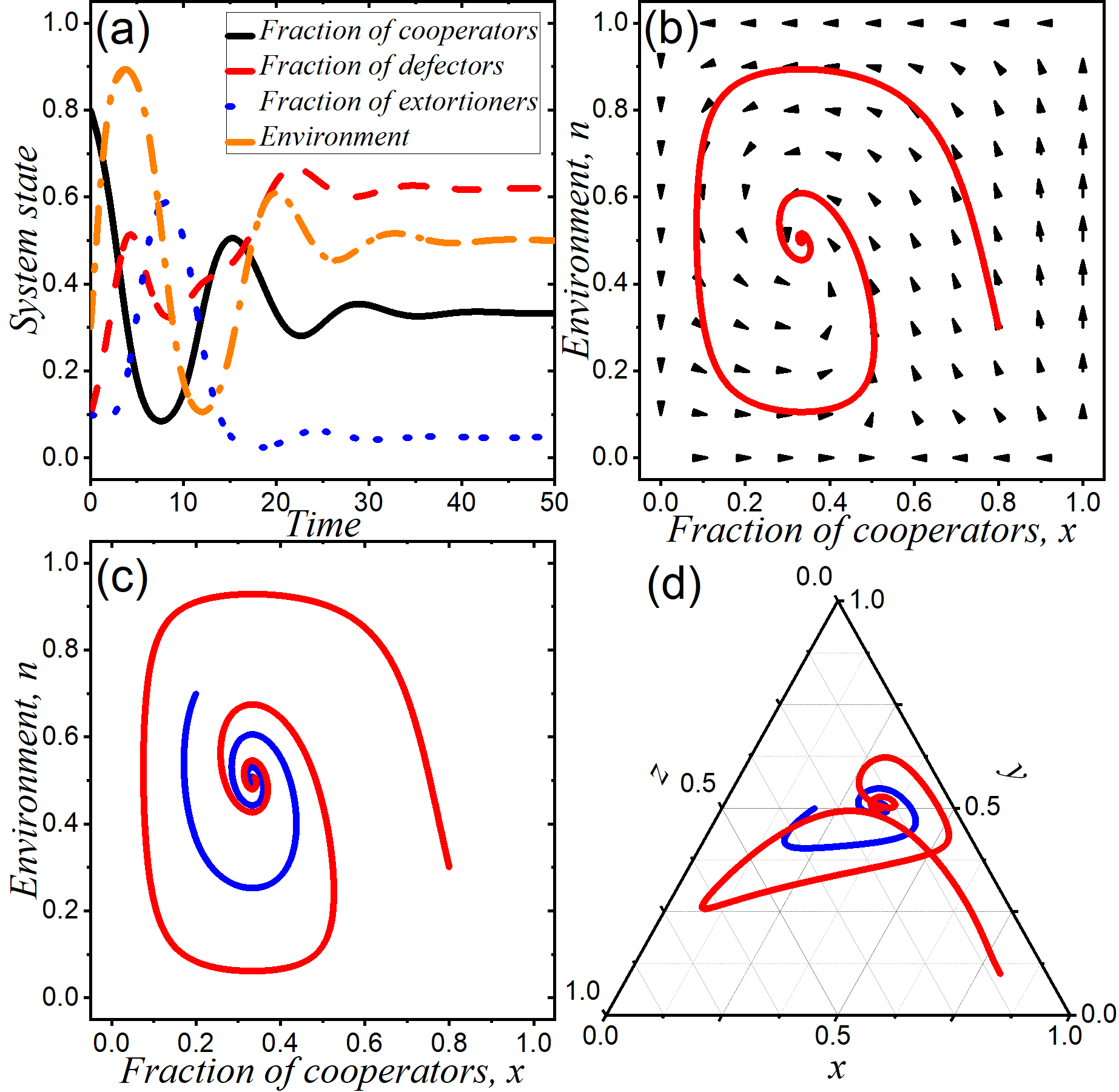}\\
  \caption{(Color online) (a) Time series of the fraction of cooperators $x$ (black), defectors $y$ (red), loners $z$ (blue) and the environmental state $n$ (orange), correspond to \textbf{Case 3} with $\varepsilon = 0.3$, $\theta = 2$, $b = 1$, $c = 0.25$ and $\delta = 0.6$.
  (b) Phase plane dynamics of $x-n$ system.  The red curve denotes the realized orbit. The black arrows denote the manifolds. The initial state is (0.8, 0.1, 0.3) and the final stable state is $(\frac{1}{1+\theta},\frac{b+c-\delta}{(1+\theta)(\delta-c)},\frac{1}{2})$.
  (c) Phase plane dynamics of $x-n$ system. (d) Phase plane dynamics of $x-y-z$ system.
  The blue and red curves represent trajectories with initial values of (0.2, 0.5, 0.7) and (0.8, 0.1, 0.3), respectively.	}
  \label{fig.3}
\end{figure}

Figure~\ref{fig.3} shows an example corresponding to \textbf{Case 3}.
Unlike \textbf{Case 1} and \textbf{Case 2}, the loneliness strategy did not become extinct, but the coexistence of all three strategies, although the frequency of loneliness was very small.
The frequency of cooperators, defectors and loners is $\frac{1}{1+\theta}$, $\frac{1+c-\delta}{(1+\theta)(\delta-c)}$ and $1-\frac{1}{(1+\theta)(\delta-c)}$, respectively.
As can be seen from $x-n$ phase diagram in Fig.~\ref{fig.3}(b), the system spirals counterclockwise towards the final stable state.

Due to $\frac{1}{1+\theta}+c < \delta < 1$, the interaction between loneliness strategy and environment becomes even greater.
Then, loners will invade an environment dominated by defectors, and the fraction of loners began to increase, and even increase to the majority of the population.
The biggest benefit brought by the increased proportion of loners is that cooperators can be further expanded.
As cooperators enhance the environment, the environmental state began to rise.
Although the loners will inevitably begin to decline, they will not go extinct in this case.
Finally, cooperators, defectors and loners coexist in the system.

Furthermore, the environmental state is neither a state of exhaustion nor a state of high consumption, but an intermediate state of replete and depleted, and the final environmental state is $\frac{1}{2}$.

For different initial states, the final stable state of the system is consistent, as shown in Fig.~\ref{fig.3}(c) and (d).
It is clear from the $x-y-z$ phase plane in Fig.\ref{fig.3}(d) that a mixed population of cooperators, defectors and loners.

Just as the rock-paper-scissors game is a paradigmatic model for biodiversity, with applications ranging from microbial populations to human societies \cite{Kerr2002,Reichenbach2007,Verma2015,Szolnoki2016}. This model, in which includes environmental feedback the biodiversity and stability of complex systems will be maintained, is therefore more representative.
These findings have important implications for maintenance and temporal development of ecological systems.

\subsection{Persistent oscillations of strategies and the environment}

\textbf{Case 4:} If $\delta +c> b$ and $y=0$, the internal equilibria $\left(\frac{1}{1+\theta},0,\frac{1}{2}\right)$ is center.

Note that centers are neutrally stable, since nearby trajectories are neither attracted to nor repelled.

\begin{figure}[ht]
  \centering
  \includegraphics[width=0.47\textwidth,clip=false]{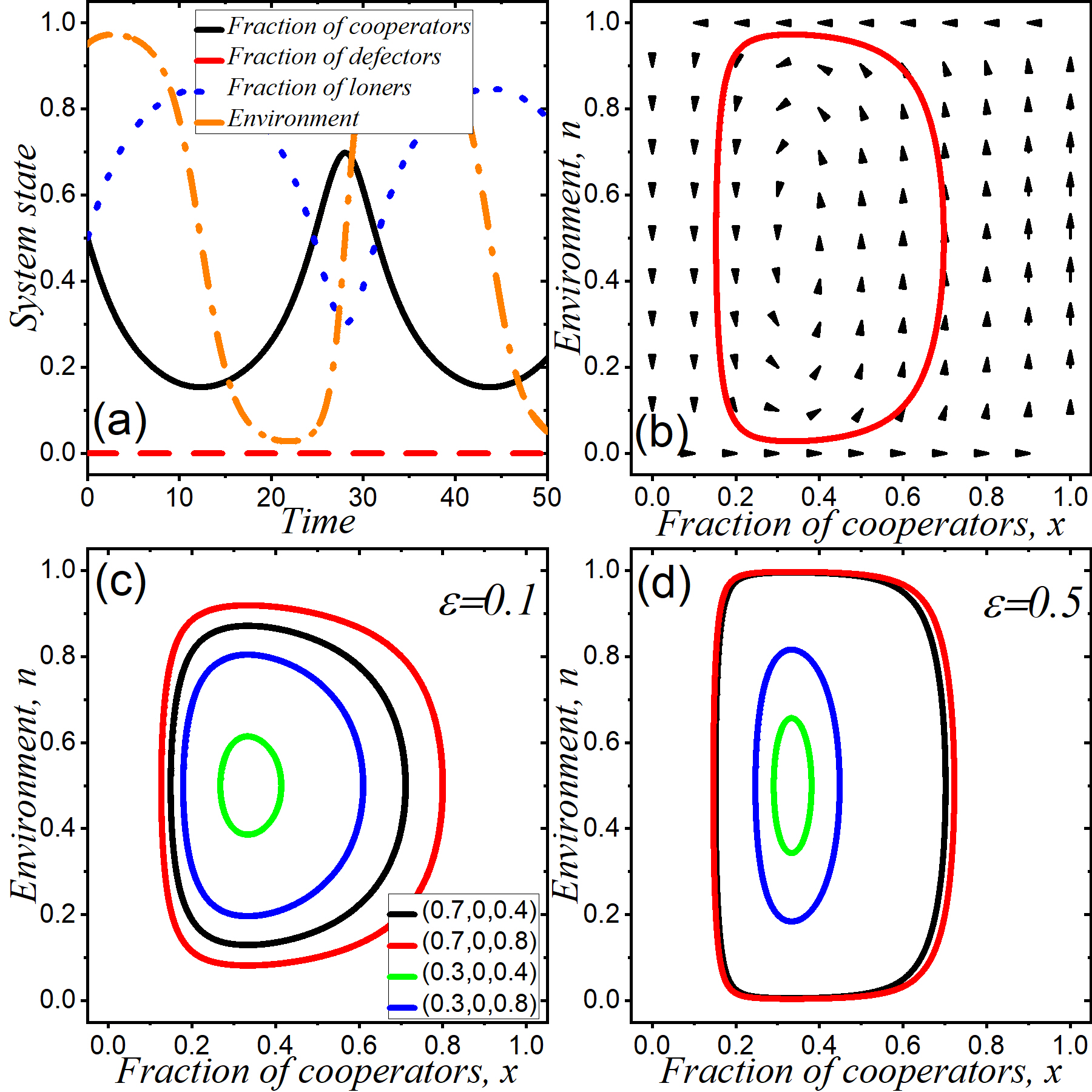}\\
  \caption{(Color online) Persistent oscillations of strategies and the environment.
  (a) Time series of the fraction of cooperators $x$ (black), defectors $y$ (red), loners $z$ (blue) and the environmental state $n$ (orange), correspond to \textbf{Case 4} with $\varepsilon = 0.3$, $\theta = 2$, $b = 1$, $c = 0.5$ and $\delta = 0.7$.
  (b) Phase plane dynamics of $x-n$ system. The red curve denotes the realized orbit. The black arrows denote the manifolds.
  The initial state is (0.5, 0, 0.95) and the interior fixed point $(\frac{1}{1+\theta},0,\frac{1}{2})$ is center.
  (c) $\varepsilon = 0.1$ and (d) $\varepsilon = 0.5$.
  Parameters are $\theta = 2$, $b = 1$, $c = 0.7$ and $\delta = 0.5$.
  The black, red, green and blue curves represent trajectories with initial values of (0.7, 0, 0.4), (0.7, 0, 0.8), (0.3, 0, 0.4) and (0.3, 0, 0.8), respectively.}
  \label{fig.4}
\end{figure}

Figure~\ref{fig.4} shows an example corresponding to the persistent oscillations of strategies and environment (\textbf{Case 4}).
The system starts from the initial state $(0.5,0,0.95)$ and then enters the state of persistent oscillations of strategies and environment.
The corresponding $x-n$ phase diagram is shown in Fig.~\ref{fig.4}(b), the state of persistent oscillations between the strategies and the environment is shown as a circle on the phase diagram.
The center of the circle is $(\frac{1}{1+\theta},0,\frac{1}{2})$.
In addition, the results are extremely robust and do not depend on $\delta>c$ or $\delta<c$.

Note that $y=0$ must be required in order to enter a state of persistent oscillation between the strategies and the environment, even at the initial moment, i.e., there is no defective strategy in the system at all times, as shown in Fig.~\ref{fig.4}(a).
This is actually back to the original two-strategy model proposed by Weitz \emph{et al}. \cite{Weitz2016}

For different initial states, as shown in Fig.~\ref{fig.4}(c) and (d), the final stable state of the
system is consistent.
The black, red, green and blue curves represent trajectories with initial values of (0.7, 0, 0.4), (0.7, 0, 0.8), (0.3, 0, 0.4) and (0.3, 0, 0.8), respectively.

It can be seen from Fig.\ref{fig.4}(c) and (d) that the relative feedback speed $\varepsilon$ affects the period of the strategy and the environment oscillation or the size of the ring in the phase diagram, but does not affect the position of the center.
A greater relative feedback speed $\varepsilon$ results in a greater impact on the environment, resulting in a steeper trajectory on the phase diagram.
The qualitative outcomes do not depend on the relative feedback speed.
This effect of the relative feedback speed is consistent in all cases.

\subsection{Interior saddle points}
\textbf{Case 5:} If $\delta +c< b$ and $y=0$, the internal equilibria $\left(\frac{1}{1+\theta},0,\frac{1}{2}\right)$ is saddle point.

\begin{figure}[ht]
  \centering
  \includegraphics[width=0.4\textwidth,clip=false]{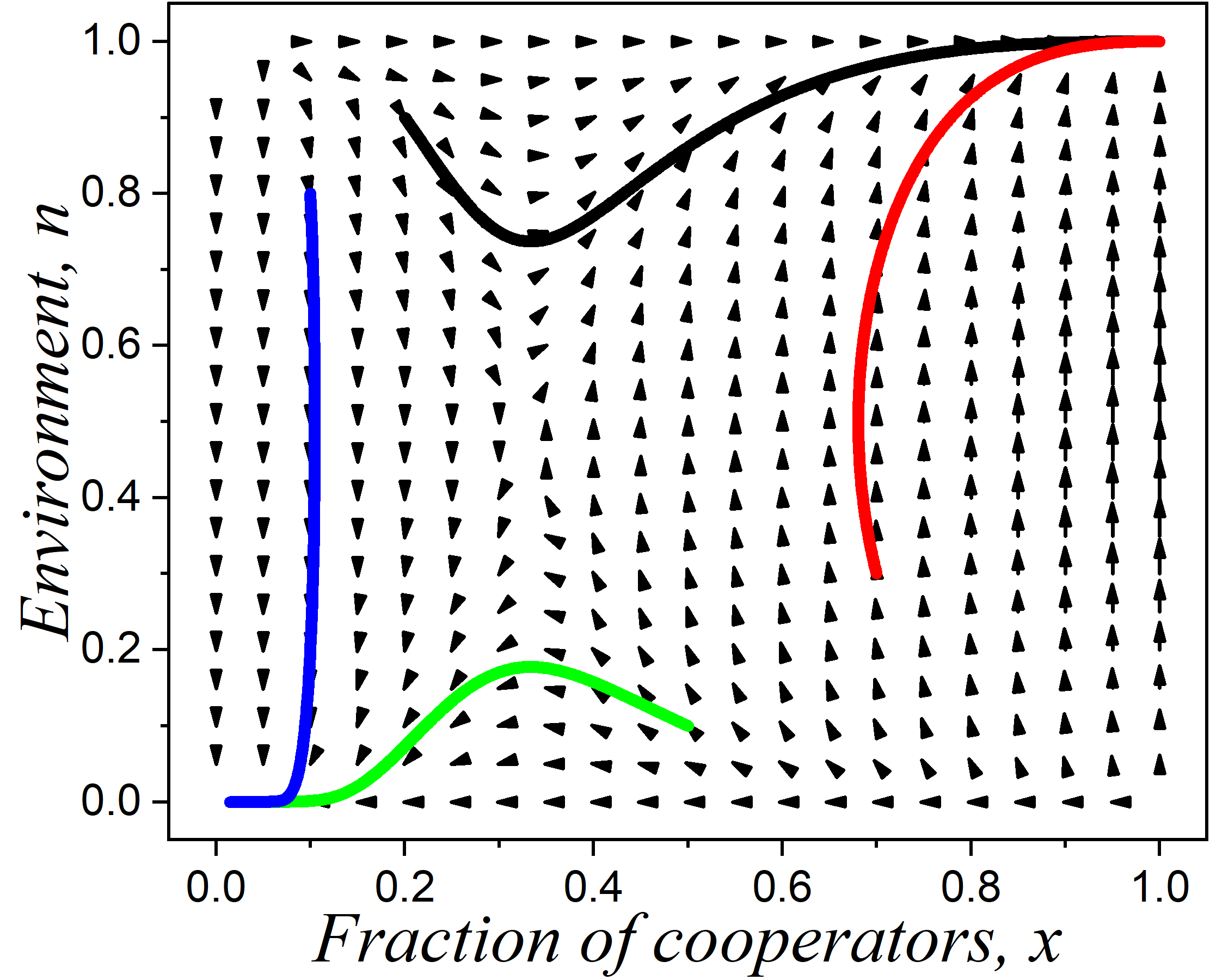}\\
  \caption{(Color online) Interior saddle points in the phase plane of strategies and the environment.
  Parameters are $\varepsilon = 0.1$, $\theta = 2$, $b = 1$, $c = 0.3$, $\delta = 0.3$ and $y=0$, which correspond to \textbf{Case 5}.
  The black arrows denote the manifolds.
  The black, red, green and blue curves represent trajectories with initial values of (0.2, 0, 0.9), (0.7, 0, 0.3), (0.5, 0, 0.1) and (0.1, 0, 0.8), respectively.}
  \label{fig.5}
\end{figure}

In Figure~\ref{fig.5}, we show the phase plane dynamics of a typical case, which correspond to \textbf{Case 5}, the internal equilibria $(\frac{1}{1+\theta},0,\frac{1}{2})$ is saddle point.
For different initial values, the cooperation and the environment co-evolve toward either ($x^{*}=1, n^{*}=1$) or ($x^{*}=0, n^{*}=0$).

The intuitive understanding of this dynamical character on the $x-n$ phase diagram can be explained as follows.
The eventual evolutionary direction of this system actually depends on the results of two types of ``competition":
(i) In the horizontal direction, it is the competition of cooperative strategy extinction and cooperative strategy dominance;
(ii) In the vertical direction there is competition between the state of environmental depleted and the state of environmental replete.

\begin{figure*}[ht]
  \centering
  \includegraphics[width=0.95\textwidth,clip=false]{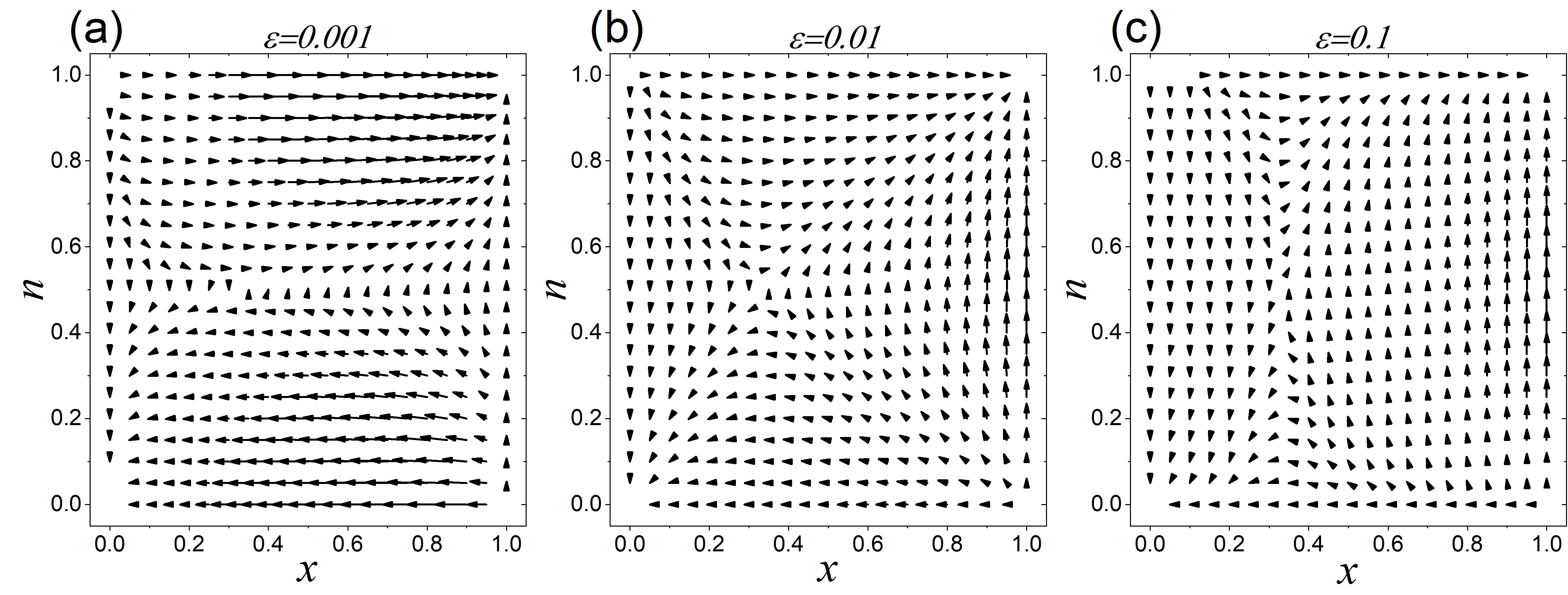}\\
  \caption{(Color online) Interior saddle points in the phase plane of strategies and the environment for different relative feedback speed $\varepsilon$.
  (a) $\varepsilon = 0.001$; (b) $\varepsilon = 0.01$; (c) $\varepsilon = 0.1$;
  The black arrows denote the manifolds.
  Parameters are $\theta = 2$, $b = 1$, $c = 0.3$, $\delta = 0.3$ and $y=0$.
  }
  \label{fig.6}
\end{figure*}

We also explore the influence of relative feedback speed as shown in Fig.~\ref{fig.6}.
We find that the relative feedback speed does not affect the position of the internal saddle point, but affects the size of the attraction domain of the two stable states.
A greater relative feedback speed $\varepsilon$ results in a greater impact on the environment, resulting in a steeper trajectory on the phase diagram.
This effect of relative feedback speed is consistent with the results in reference~\cite{Liu2021}.

\section{Conclusions}
In this paper, we have extended the classical two-strategy model to three-strategy model. For general return-dependent feedback dynamics, we have given the conditions for the existence and stability of internal equilibrium respectively. We find that there are two-strategy coexistence states, three-strategy coexistence states, persistent oscillations of strategy and environment, and internal saddle points in this system.
These states are determined by the relative feedback strength and payoff matrix, and are independent of the relative feedback speed and initial state.

There are two types of two-strategy coexistence (coexistence of cooperators and defectors, loners are become extinction).
The first type is that the environmental state is in the state of depleted, and the second type is that the environmental state of high consumption.

In the coexistence of three strategies, when the evolution of the system is stable, different from the above two-strategy coexistence, the loneliness strategy is not extinct, but the three strategies of cooperation, defection and loneliness coexist, although the frequency of loneliness was small. And the environment state is an intermediate state between replete and depleted.
The system spirals counterclockwise towards the final stable state on the phase diagram.

In particular, the three-strategy coexistence provides a new mechanism for maintaining biodiversity in biology, ecology, and sociology.
Biodiversity is essential to the viability of ecological systems.
One of the central aims of ecology is to identify mechanisms that maintain biodiversity.
However, previous research mechanisms seldom consider environmental feedback. Therefore, this model which considering environmental feedback is more promising in biology, ecology, or sociology.

When there is no defective strategy at the initial moment, the persistent oscillation results of our three-strategy model are similar to those of the classical two-strategy model. That is, when $y=0$, the three-strategy model returns to the two-strategy model.
The greater relative feedback speed results in a greater impact on the environment, resulting in a steeper trajectory on the phase diagram, but does not affect the position of the center.
The qualitative outcomes do not depend on the relative feedback speed.
This effect of relative feedback speed is consistent with the results of the classical two-strategy model proposed by Weitz ~\cite{Weitz2016}.

We also showed that the internal saddle points on the phase diagram, the system starts from different initial states, and may eventually go to different stable states $(0,0)$ or $(1,1)$. The two stable states have different domains of attraction and are affected by the relative feedback speed.

In addition, the biggest difference between the results of this three-strategy model and the original two-strategy model is that limit cycle and heteroclinic cycle trajectory are not found. Why it doesn't exist requires further research.
Also, most current models considering environmental feedback focus on well-mixed and infinite populations, while more realistic finite population models with spatial structure are worthy of further study.

\section*{Acknowledgments}
This work was supported by the National Natural Science Foundation of China (Grants No. 11975111 and No. 12047501).

\appendix*
\section{Stability of interior fixed points}
\label{appendix}

Differential equations describing the evolutionary system can be written as
\begin{equation}
\label{eq.a1}
\left\{
  \begin{array}{ll}
   \dot{x}=x \left[ (r_{C}-r_{L})(1-x)-(r_{D}-r_{L})y \right], \\
   \dot{y}=y \left[ (r_{D}-r_{L})(1-y)-(r_{C}-r_{L})x \right], \\
   \dot{n}=\varepsilon n(1-n) \left[ (\theta+1)x-1 \right],
  \end{array}
\right.
\end{equation}
where
\begin{equation}
\label{eq.a2}
   r_{C}-r_{L}=(1-2n)\left[ (\delta+c-b)x+(\delta+c)y \right] ,
\end{equation}
and
\begin{equation}
\label{eq.a3}
   r_{D}-r_{L}=\left[ c-n(\delta+c-b) \right] x + \left[ c-n(\delta+c)\right] y.
\end{equation}

Jacobian of this system is
\begin{equation}
\label{eq.a4}
  J=\left[
  \begin{array}{ccc}
    J_{11} & J_{12} & J_{13} \\
    J_{21} & J_{22} & J_{23} \\
    \varepsilon n(1-n)(1+\theta) & 0 & \varepsilon(1-2n)[(\theta+1)x-1] \\
  \end{array}
\right]
\end{equation}
where
\begin{equation}
\label{eq.a5}
\begin{split}
  J_{11}&= 3 b x^2 - 3 c x^2 - 6 b n x^2 + 6 c n x^2 - 3 x^2 \delta   \\
     & + 6 n x^2 \delta + n y^2 \delta  - c y^2 +c n y^2 - 4 c x y    \\
     & - 2 b n x y + 6 c n x y + 6 n x y \delta - 2 x y \delta     \\
     & + 4 b n x - 4 c n x + 2 c x -2 b x - 2 c n y \\
     & + c y + y \delta - 2 n y \delta + 2 x \delta - 4 n x \delta  ,
\end{split}
\end{equation}
\begin{equation}
\label{eq.a6}
\begin{split}
  J_{12}=&x\{[3n(\delta+c)-2c-nb-\delta]x  \\
       &+2[n(\delta+c)-c]y-(2n-1)(\delta+c) \},
\end{split}
\end{equation}
\begin{equation}
\label{eq.a7}
  J_{13}=x(2-2x-y)\{[b-(\delta+c)]x-(\delta+c)y \},
\end{equation}
\begin{equation}
\label{eq.a8}
\begin{split}
  J_{21}=& y \{ [n (4 x + 3 y -1)- 2 x - y] \delta \\
        &+ c [ 1 - 2 x - 2 y + n (-1 + 4 x + 3 y)]   \\
        &- b n (-1 + 4 x + y) + 2 b x \},
\end{split}
\end{equation}
\begin{equation}
\label{eq.a9}
\begin{split}
  J_{22}=&(2n-1)\left( \delta+c-b \right)x^{2} + 2(2n-1)(\delta+c)xy  \\
        &+[n(\delta+c-b)-c](2y-1)x     \\
        &+[c-n(\delta+c)](2-3y)y,
\end{split}
\end{equation}
and
\begin{equation}
\label{eq.a10}
  J_{23}=y(2x+y-1)\{(\delta+c-b)x+(\delta+c)y\}.
\end{equation}

Without loss of generality, we set $b = 1$ to analyze the stability of these interior fixed points.

\section*{Case 1}

The interior fixed point is $\left( x^{*}=\delta,y^{*}=1-\delta,n^{*}=0\right) $.
The Jacobian of this interior fixed point is:
\begin{equation}
\label{eq.a11}
  J=
\left[
  \begin{array}{ccc}
    \delta(2\delta-1-c-\delta^2) & \delta^{2}-\delta^{3}-c\delta & -c\delta(1-\delta) \\
    (1-\delta)(\delta-\delta^{2}-c) & (\delta^{2}+c)(\delta-1) & c\delta(1-\delta) \\
    0 & 0 & \varepsilon(\theta+1)\delta-\varepsilon \\
  \end{array}
\right].
\end{equation}
The eigen equation is
\begin{equation}
\label{eq.a12}
\begin{split}
  |J-\lambda E| = & \{ [\delta(2\delta-1-c-\delta^2)-\lambda][(\delta^{2}+c)(\delta-1)-\lambda] \\ &-(\delta^{2}-\delta^{3}-c\delta)(1-\delta)(\delta-\delta^{2}-c) \}  \\
  & \times \{\varepsilon[(\theta+1)\delta-1]-\lambda \},
\end{split}
\end{equation}
that is
\begin{equation}
\label{eq.a13}
  |J-\lambda E| = (c+\lambda)[\delta(1-\delta)+\lambda] \{\varepsilon[(\theta+1)\delta-1]-\lambda \}.
\end{equation}
We can derive eigenvalues are
\begin{equation}
\label{eq.a14}
\begin{split}
   & \lambda_{1}=-c, \\
   & \lambda_{2} =-\delta(1-\delta) ,  \\
   & \lambda_{3} = \varepsilon[(\theta+1)\delta-1].
\end{split}
\end{equation}
Because $0<c<1$, $0<\delta<1$ and $\varepsilon >0$, then $\lambda_{1} <0$ and $\lambda_{2} <0$.
Thus, if $\delta < \frac{1}{1+\theta}$, eigenvalue $\lambda_{3} <0$ and the fixed point is stable equilibrium.

\section*{Case 2}

The interior fixed point is $( x^{*}=\frac{1}{1+\theta},y^{*}=\frac{\theta}{1+\theta},n^{*}=\frac{\delta+\theta \delta-1}{(1+\theta)(\delta+c)-1}) $.
The Jacobian of this interior fixed point is:

\begin{equation}
\label{eq.a15}
  J=
\left[
  \begin{array}{ccc}
    J_{(11)} & J_{(12)} & J_{(13)} \\
    J_{(21)} & J_{(22)} & J_{(23)} \\
    J_{(31)} & 0 & 0 \\
  \end{array}
\right],
\end{equation}
where
\begin{equation}
\label{eq.a16}
  J_{(11)} =\frac{-c \theta^2 - c^2 (1 + \theta)^2 + ( \delta + \delta \theta -1 )^2}
  {(1+\theta)^2 [(1+\theta)(\delta+c)-1]},
\end{equation}
\begin{equation}
\label{eq.a17}
J_{(12)}=\frac{c\theta-c^{2}(1+\theta)^{2}+(\delta+\delta\theta-1)^{2}}{(1+\theta)^{2} [(1+\theta)(\delta+c)-1]},
\end{equation}
\begin{equation}
\label{eq.a18}
J_{(13)}=\frac{\theta}{(1+\theta)^{2}} \left( \frac{1}{1+\theta}-\delta-c \right) ,
\end{equation}
\begin{equation}
\label{eq.a19}
J_{(21)}=\theta \frac{ c \theta - c^2 (1 + \theta)^2 + ( \delta + \delta \theta -1 )^2}
{(1+\theta)^2 [(1+\theta)(\delta+c)-1]},
\end{equation}
\begin{equation}
\label{eq.a20}
J_{(22)}=\theta \frac{(\delta+\delta\theta-1)^{2}-c^2 (1+\theta)^2 -c}
{(1+\theta)^2 [(1+\theta)(\delta+c)-1]},
\end{equation}
\begin{equation}
\label{eq.a21}
J_{(23)}=-\frac{\theta}{(1+\theta)^{2}} \left( \frac{1}{1+\theta}-\delta-c \right) ,
\end{equation}
and
\begin{equation}
\label{eq.a22}
J_{(31)}=\frac{ c \varepsilon (\delta+\delta\theta-1)(1+\theta)^2}{[(1+\theta)(\delta+c)-1]^2}.
\end{equation}
The eigen equation is
\begin{equation}
\label{eq.a23}
\begin{split}
    |J-\lambda E| = & -\lambda(J_{(11)}-\lambda)(J_{(22)}-\lambda) \\
    & -J_{(13)}J_{(31)}(J_{(22)}-\lambda) \\
    & +J_{(12)}J_{(23)}J_{(31)} + \lambda J_{(12)}J_{(21)},
\end{split}
\end{equation}
that is
\begin{equation}
\label{eq.a24}
\begin{split}
  |J-\lambda E| = & \lambda(J_{(13)}J_{(31)} + J_{(12)}J_{(21)}-J_{(11)}J_{(22)})  \\
           & - J_{(13)}J_{(31)}(J_{(12)}+J_{(22)}) \\
         & + \lambda^2 (J_{(11)}+J_{(22)})-\lambda^3 .
\end{split}
\end{equation}
We can derive eigenvalues are
\begin{equation}
\label{eq.a25}
\begin{split}
   & \lambda_{1} = \frac{(1+\theta)(\delta-c)-1}{1+\theta}, \\
   & \lambda_{2} = \frac{-c \theta - \sqrt{\alpha}}
     {2 \beta},  \\
   & \lambda_{3} =\frac{-c \theta + \sqrt{\alpha}}
     {2 \beta}.
\end{split}
\end{equation}
where
\begin{equation}
\label{eq.a26}
\begin{split}
  & \alpha= c^2 \theta^2 - 4 \beta \left( -c \varepsilon \theta + c \delta \varepsilon \theta + c \delta \varepsilon \theta^2 \right) , \\
 & \beta = c -1 + \delta - \theta + 2 c \theta + 2\delta \theta + c \theta^2 + \delta \theta^2 .
\end{split}
\end{equation}
If eigenvalue $\lambda_{1} <0$ that
\begin{equation}
\label{eq.a27}
   \delta < \frac{1}{1+\theta}+c.
\end{equation}
If eigenvalue $\lambda_{2} <0$ and $\lambda_{3} <0$ that
\begin{equation}
\label{eq.a28}
\left\{
  \begin{array}{ll}
 \alpha > 0 \\
   -1 + c + \delta - \theta + 2 c \theta + 2 \delta \theta + c \theta^2 + \delta \theta^2 > 0.
  \end{array}
\right.
\end{equation}
Because $0<c<1$, $0<\delta<1$ and $\varepsilon >0$, then Eq.(\ref{eq.a28}) can be written as
\begin{equation}
\label{eq.a29}
\left\{
  \begin{array}{lll}
   \frac{1}{1+\theta}-\frac{c}{2}-\frac{1}{2} \sqrt{\frac{c\theta}{\varepsilon (1+\theta)^3 }+c^2} < \delta \\
   \delta < \frac{1}{1+\theta}-\frac{c}{2}+\frac{1}{2} \sqrt{\frac{c\theta}{\varepsilon (1+\theta)^3 }+c^2} \\
    \delta >  \frac{1}{1+\theta}-c,
  \end{array}
\right.
\end{equation}
that is
\begin{equation}
\label{eq.a30}
   \frac{1}{1+\theta}-c < \delta < \frac{1}{1+\theta}-\frac{c}{2}+\frac{1}{2} \sqrt{\frac{c\theta}{\varepsilon (1+\theta)^3 }+c^2}.
\end{equation}
According to $0 < n^{*} < 1$, we have $\delta > \frac{1}{1+\theta}$.
Thus, when $\varepsilon$ very small, if $\frac{1}{1+\theta} < \delta < \frac{1}{1+\theta}+c$, eigenvalues $\lambda_{1,2,3} <0$, and the fixed point is stable equilibrium.

\section*{Case 3}

The interior fixed point is ($x^{*}=\frac{1}{1+\theta},y^{*}=\frac{\delta-c-1}{(1+\theta)(c-\delta)},n^{*}=\frac{1}{2}$).
The Jacobian of this interior fixed point is:
\begin{equation}
\label{eq.a31}
  J=
\left[
  \begin{array}{ccc}
    J_{(11)} & J_{(12)} & J_{(13)} \\
    J_{(21)} & J_{(22)} & J_{(23)} \\
    J_{(13)} & 0 & 0 \\
  \end{array}
\right],
\end{equation}
where
\begin{equation}
\label{eq.a32}
  J_{(11)} = \frac{(1+c-\delta)^2}{2(c-\delta)(1+\theta)^{2}},
\end{equation}
\begin{equation}
\label{eq.a33}
  J_{(12)} = \frac{1+c-\delta}{2(1+\theta)^{2}},
\end{equation}
\begin{equation}
\label{eq.a34}
  J_{(13)} = \frac{(2\theta+1)(c-\delta)+1}{(c-\delta)^{2}(1+\theta)^{2}}\frac{2c}{1+\theta},
\end{equation}
\begin{equation}
\label{eq.a35}
  J_{(21)} = \frac{(1+c-\delta)^2 ((2+\theta)(\delta-c)-1)}{2(c-\delta)^2 (1+\theta)^2},
\end{equation}
\begin{equation}
\label{eq.a36}
  J_{(22)} = \frac{(\delta-c-1)(1+(2+\theta)(c-\delta))}{2(c-\delta)(1+\theta)^2},
\end{equation}
\begin{equation}
\label{eq.a37}
  J_{(23)} = \frac{2c(\delta-c-1)(1+c\theta-\theta\delta)}{(c-\delta)^3 (1+\theta)^3 },
\end{equation}
and
\begin{equation}
\label{eq.a38}
  J_{(13)} = \frac{\varepsilon(1+\theta)}{4}.
\end{equation}
The eigen equation is
\begin{equation}
\label{eq.a39}
\begin{split}
  |J-\lambda E| = & \lambda^2 (J_{(11)}+J_{(22)})-\lambda^3+\lambda(J_{(13)}J_{(31)}  \\
    & + J_{(12)}J_{(21)}-J_{(11)}J_{(22)})  \\
    & + J_{(31)}(J_{(12)}J_{(23)}-J_{(13)}J_{(22)}).
\end{split}
\end{equation}

According to $0<c<1$, $0<\delta<1$, $\theta>0$ and $\varepsilon >0$, we can derive eigenvalues are $\lambda_{1,2,3} <0$.

Because $0 < y^{*} < 1$, we have $c< \delta < 1+c$.

Thus, if $\frac{1}{1+\theta} + c< \delta < 1$ and the fixed point is stable equilibrium.

\section*{Case 4}

The interior fixed point is $(x^{*}=\frac{1}{1+\theta},y^{*}=0,n^{*}=\frac{1}{2})$.
The Jacobian of this interior fixed point is:
\begin{equation}
\label{eq.a40}
  J=
\left[
  \begin{array}{ccc}
    0 & \frac{\delta-c-1}{2(1+\theta)^{2}} & \frac{2\theta(1-\delta-c)}{(1+\theta)^{3}} \\
    0 & \frac{c+1-\delta}{2(1+\theta)} & 0 \\
    \frac{\varepsilon(1+\theta)}{4} & 0 & 0 \\
  \end{array}
\right].
\end{equation}
The eigen equation is
\begin{equation}
\label{eq.a41}
    |J-\lambda E| = \left(\frac{c+1-\delta}{2(1+\theta)}-\lambda \right) \left(\lambda^2 - \frac{\varepsilon\theta(1-\delta-c)}{2(1+\theta)^{2}} \right).
\end{equation}
We can derive eigenvalues are
\begin{equation}
\label{eq.a42}
    \lambda_{1} = \frac{c+1-\delta}{2(1+\theta)}, \lambda_{2,3}=\pm \sqrt{\frac{\varepsilon\theta(1-\delta-c)}{2(1+\theta)^{2}}}.
\end{equation}
According to $0<c<1$, $0<\delta<1$, $\theta>0$ and $\varepsilon >0$, eigenvalue $\lambda_{1} >0$.
And, if $1 < \delta+c$, the eigenvalues $\lambda_{2,3}$ are complex.
So the fixed point is center.

\section*{Case 5}

According to \textbf{CASE 4}, if $1 > \delta+c$, then one of the two eigenvalues $\lambda_{2,3}$ is positive while the other is negative.
Thus, the fixed point $(x^{*}=\frac{1}{1+\theta},y^{*}=0,n^{*}=\frac{1}{2})$ is a saddle point.

\bibliography{optional-participation-lbq}
\end{document}